

\def\singlespace{\normalbaselines}
\def\oneandahalfspace{\baselineskip=1.15\normalbaselineskip plus 1pt
\lineskip=2pt\lineskiplimit=1pt}

\def\np{\vfill\eject}

\def\nofirstpagenoten{\nopagenumbers\footline={\ifnum\pageno>1\tenrm
\hss\folio\hss\fi}}
\def\nofirstpagenotwelve{\nopagenumbers\footline={\ifnum\pageno>1\twelverm
\hss\folio\hss\fi}}
\def\leaderfill{\leaders\hbox to 1em{\hss.\hss}\hfill}
\def\ft#1#2{{\textstyle{{#1}\over{#2}}}}
\def\frac#1/#2{\leavevmode\kern.1em
\raise.5ex\hbox{\the\scriptfont0 #1}\kern-.1em/\kern-.15em
\lower.25ex\hbox{\the\scriptfont0 #2}}
\def\sfrac#1/#2{\leavevmode\kern.1em
\raise.5ex\hbox{\the\scriptscriptfont0 #1}\kern-.1em/\kern-.15em
\lower.25ex\hbox{\the\scriptscriptfont0 #2}}


\parindent=20pt
\def\narrow{\advance\leftskip by 40pt \advance\rightskip by 40pt}

\def\AB{\bigskip
        \centerline{\bf ABSTRACT}\medskip\narrow}
\def\nonarrower{\advance\leftskip by -40pt\advance\rightskip by -40pt}
\def\AE{\bigskip\nonarrower}

\def\boxit#1{\vbox{\hrule\hbox{\vrule\kern3pt
        \vbox{\kern3pt#1\kern3pt}\kern3pt\vrule}\hrule}}

\def\gtorder{\mathrel{\raise.3ex\hbox{$>$}\mkern-14mu
             \lower0.6ex\hbox{$\sim$}}}
\def\ltorder{\mathrel{\raise.3ex\hbox{$<$}|mkern-14mu
             \lower0.6ex\hbox{\sim$}}}
\def\dalemb#1#2{{\vbox{\hrule height .#2pt
        \hbox{\vrule width.#2pt height#1pt \kern#1pt
                \vrule width.#2pt}
        \hrule height.#2pt}}}

\font\fourteentt=cmtt10 scaled \magstep2
\font\fourteenbf=cmbx12 scaled \magstep1
\font\fourteenrm=cmr12 scaled \magstep1
\font\fourteeni=cmmi12 scaled \magstep1
\font\fourteenss=cmss12 scaled \magstep1
\font\fourteensy=cmsy10 scaled \magstep2
\font\fourteensl=cmsl12 scaled \magstep1
\font\fourteenex=cmex10 scaled \magstep2
\font\fourteenit=cmti12 scaled \magstep1
\font\twelvett=cmtt10 scaled \magstep1 \font\twelvebf=cmbx12
\font\twelverm=cmr12 \font\twelvei=cmmi12
\font\twelvess=cmss12 \font\twelvesy=cmsy10 scaled \magstep1
\font\twelvesl=cmsl12 \font\twelveex=cmex10 scaled \magstep1
\font\twelveit=cmti12
\font\tenss=cmss10
 
 \font\ninebf=cmbx7 scaled \magstep1
\font\ninerm=cmr7 scaled \magstep1 \font\ninei=cmmi7 scaled \magstep1
\font\ninesy=cmsy7 scaled \magstep1 
\font\eightrm=cmr7 scaled 1140 
 
\font\sevenbf=cmbx7 \font\sevenrm=cmr7 \font\seveni=cmmi7
\font\sevensy=cmsy7 

\catcode`@=11
\newskip\ttglue
\newfam\ssfam

\def\fourteenpoint{\def\rm{\fam0\fourteenrm}
\textfont0=\fourteenrm \scriptfont0=\tenrm \scriptscriptfont0=\sevenrm
\textfont1=\fourteeni \scriptfont1=\teni \scriptscriptfont1=\seveni
\textfont2=\fourteensy \scriptfont2=\tensy \scriptscriptfont2=\sevensy
\textfont3=\fourteenex \scriptfont3=\fourteenex \scriptscriptfont3=\fourteenex
\def\it{\fam\itfam\fourteenit} \textfont\itfam=\fourteenit
\def\sl{\fam\slfam\fourteensl} \textfont\slfam=\fourteensl
\def\bf{\fam\bffam\fourteenbf} \textfont\bffam=\fourteenbf
\scriptfont\bffam=\tenbf \scriptscriptfont\bffam=\sevenbf
\def\tt{\fam\ttfam\fourteentt} \textfont\ttfam=\fourteentt
\def\ss{\fam\ssfam\fourteenss} \textfont\ssfam=\fourteenss
\tt \ttglue=.5em plus .25em minus .15em
\normalbaselineskip=16pt
\abovedisplayskip=16pt plus 4pt minus 12pt
\belowdisplayskip=16pt plus 4pt minus 12pt
\abovedisplayshortskip=0pt plus 4pt
\belowdisplayshortskip=9pt plus 4pt minus 6pt
\parskip=5pt plus 1.5pt
\setbox\strutbox=\hbox{\vrule height12pt depth5pt width0pt}
\let\sc=\tenrm
\let\big=\fourteenbig \normalbaselines\rm}
\def\fourteenbig#1{{\hbox{$\left#1\vbox to12pt{}\right.\n@space$}}}

\def\twelvepoint{\def\rm{\fam0\twelverm}
\textfont0=\twelverm \scriptfont0=\ninerm \scriptscriptfont0=\sevenrm
\textfont1=\twelvei \scriptfont1=\ninei \scriptscriptfont1=\seveni
\textfont2=\twelvesy \scriptfont2=\ninesy \scriptscriptfont2=\sevensy
\textfont3=\twelveex \scriptfont3=\twelveex \scriptscriptfont3=\twelveex
\def\it{\fam\itfam\twelveit} \textfont\itfam=\twelveit
\def\sl{\fam\slfam\twelvesl} \textfont\slfam=\twelvesl
\def\bf{\fam\bffam\twelvebf} \textfont\bffam=\twelvebf
\scriptfont\bffam=\ninebf \scriptscriptfont\bffam=\sevenbf
\def\tt{\fam\ttfam\twelvett} \textfont\ttfam=\twelvett
\def\ss{\fam\ssfam\twelvess} \textfont\ssfam=\twelvess
\tt \ttglue=.5em plus .25em minus .15em
\normalbaselineskip=14pt
\abovedisplayskip=14pt plus 3pt minus 10pt
\belowdisplayskip=14pt plus 3pt minus 10pt
\abovedisplayshortskip=0pt plus 3pt
\belowdisplayshortskip=8pt plus 3pt minus 5pt
\parskip=3pt plus 1.5pt
\setbox\strutbox=\hbox{\vrule height10pt depth4pt width0pt}
\let\sc=\ninerm
\let\big=\twelvebig \normalbaselines\rm}
\def\twelvebig#1{{\hbox{$\left#1\vbox to10pt{}\right.\n@space$}}}

\def\tenpoint{\def\rm{\fam0\tenrm}
\textfont0=\tenrm \scriptfont0=\sevenrm \scriptscriptfont0=\fiverm
\textfont1=\teni \scriptfont1=\seveni \scriptscriptfont1=\fivei
\textfont2=\tensy \scriptfont2=\sevensy \scriptscriptfont2=\fivesy
\textfont3=\tenex \scriptfont3=\tenex \scriptscriptfont3=\tenex
\def\it{\fam\itfam\tenit} \textfont\itfam=\tenit
\def\sl{\fam\slfam\tensl} \textfont\slfam=\tensl
\def\bf{\fam\bffam\tenbf} \textfont\bffam=\tenbf
\scriptfont\bffam=\sevenbf \scriptscriptfont\bffam=\fivebf
\def\tt{\fam\ttfam\tentt} \textfont\ttfam=\tentt
\def\ss{\fam\ssfam\tenss} \textfont\ssfam=\tenss
\tt \ttglue=.5em plus .25em minus .15em
\normalbaselineskip=12pt
\abovedisplayskip=12pt plus 3pt minus 9pt
\belowdisplayskip=12pt plus 3pt minus 9pt
\abovedisplayshortskip=0pt plus 3pt
\belowdisplayshortskip=7pt plus 3pt minus 4pt
\parskip=0.0pt plus 1.0pt
\setbox\strutbox=\hbox{\vrule height8.5pt depth3.5pt width0pt}
\let\sc=\eightrm
\let\big=\tenbig \normalbaselines\rm}
\def\tenbig#1{{\hbox{$\left#1\vbox to8.5pt{}\right.\n@space$}}}
\let\rawfootnote=\footnote \def\footnote#1#2{{\rm\parskip=0pt\rawfootnote{#1}
{#2\hfill\vrule height 0pt depth 6pt width 0pt}}}

\def\tenfoot{\tenpoint\hskip-\parindent\hskip-.1cm}

\overfullrule=0pt
\twelvepoint
\oneandahalfspace
\def\sbullet{\raise.2em\hbox{$\scriptscriptstyle\bullet$}}
\nofirstpagenotwelve
\hsize=16.5 truecm
\baselineskip 15pt

\def\ft#1#2{{\textstyle{{#1}\over{#2}}}}

\def\lra{\leftrightarrow}

\def\ul{\underline}

\oneandahalfspace
\rightline{CTP TAMU--67/93}
\rightline{UG-5/93}
\rightline{hep-th/9312168}
\rightline{December 1993}

\vskip 2truecm
\centerline{\bf Twistor-like Formulation of Super $p$--Branes}
\vskip 1.5truecm
\centerline{E. Bergshoeff $^1$ and E. Sezgin $^2$
\footnote{$^\dagger$}{\tenfoot \sl  Supported in part by the National
Science Foundation, under grant PHY-9106593.}}
\vskip 1.5truecm
\noindent{$^1$ {\sl Institute for Theoretical Physics,
Nijenborgh 4, 9747 AG Groningen, The Netherlands.}}
\vskip 0.5truecm
\noindent{$^2${\sl  Center for Theoretical Physics,
Texas A\&M University, College Station, TX 77843--4242, USA.}}
\vskip 1.5truecm
\AB\singlespace

 Closed super $(p+2)$--forms in target superspace are relevant for the
construction of the usual super $p$--brane actions. Here we construct closed
super $(p+1)$--forms on a {\it worldvolume superspace}. They are built out of
the pull-backs  of the Kalb-Ramond super $(p+1)$--form and its curvature. We
propose a twistor-like formulation of a class of super $p$--branes which
crucially depends on the existence of these closed super $(p+1)$--forms.

\AE\oneandahalfspace

\np

\vskip 1.5truecm

\noindent{\bf 1. Introduction}
\bigskip
The manifestly spacetime supersymmetric
formulation of string theory
\`a la Green and Schwarz has a fermionic gauge symmetry, known as
the $\kappa$--symmetry [1] . This symmetry is of crucial importance for the
model, but also gives rise to formidable problems in
its quantization. A few years ago a geometrical understanding of
$\kappa$--symmetry has emerged after the work of ref.~[2]
which also holds the promise of better prospects for a
quantization of the model [3]. It should be emphasized, however, that
a fully covariant quantization scheme has not yet emerged.

In the simple situation of a superparticle in $d=3$ dimensions,
the theory was reformulated in such a way that the
$\kappa$--symmetry
can be interpreted as $N=1$ local worldline supersymmetry [2]. The key to
this formulation is the introduction of twistor-like variables, $\lambda$,
which are commuting spinors arising as the superpartners of the target
superspace fermionic coordinates
\footnote{$^\dagger$}{\tenfoot The word {\it twistor-like} is used to avoid
confusion
with the supertwistor which consists of a multiplet of fields forming a
multiplet of
superconformal groups which are known to exists in dimensions $d\le 6$. In fact
such variables have been used previously in a twistor formulation of
superparticles and superstrings in $d=3,4,6$ [4]. A similar, but not quite the
same, multiplet of variables were introduced in [5] to give a twistor-like
formulation of these models in $d=10$. The twistor--like formulation of
ref. [2], which we will be following in this paper, differs from both.}
. The idea is essentially to make the change
of variable  $P^\mu\rightarrow
\lambda^\alpha \gamma^\mu_{\alpha\beta}\lambda^\beta$ such that the mass
shell constraint $P^\mu P_\mu=0$ is satisfied, and that local supersymmetry
is now formulated with
the help of the new variable $\lambda$. This construction was
later generalized to superparticles in higher dimensions [6-10], type-I
superstrings in $d=3,4,6$ [11] and $d=10$ [12-14] ,
type-II superstrings in $d=3$ [15] and supermembranes in $d=11$ [16]. After
these works it became clear that there exists a closed super $(p+1)$--form
$(p=0,1,2)$ on the worldvolume superspace which plays a central role. This
should be contrasted with  the crucial role the super $(p+2)$--forms in
target superspace play for the existence of the usual super $p$--branes. In
fact, the closed super $(p+1)$--form on the worldvolume superspace is  built
out of the pull-backs  of the super $(p+1)$--form and its curvature in target
superspace.

The purpose of this paper is to investigate twistor-like formulations
of other the super $p$--branes as well. Beyond the cases discussed above, there
are four more cases in the usual brane-scan: $(p=3, d=6,8)$, $(p=4,d=9)$ and
$(p=5, d=10)$ where $d$ is the dimension of the target spacetime. We construct
closed super $(p+1)$--forms for $(p=3, d=6)$ and $(p=5, d=10)$  and using
these forms we  propose an action for the twistor-like formulation of these
theories, thereby  generalizing  previous results mentioned above ( whether the
obstacles encountered for the cases of $(p=3, d=8)$ and $(p=4, d=9)$ are
circumventable remains to be seen). We hope that, among other things, this
formulation will be useful in search of the so far elusive heterotic
$5$--brane action.

The case of the superstring is somewhat special due to the
extra
world-sheet Weyl symmetry. This case has been treated in great detail in
[13].  Here we shall focus on super $p$--branes
with $p\ne 1$, of which the massive superparticle ($p=0$) is
the simplest, and therefore we begin with its description. The massive
particle in $d=2$ with worldline $n=1$ local supersymmetry has been
considered in [17][18], and in $d=3$ with $n=2$, in [18]. The massive
particle action which  will be presented here has the maximal $n=8$ local
worldline supersymmetry.

 \bigskip
\noindent{\bf 2. The Massive Superparticle\ (p=0)}
\bigskip
        Consider a superspace ${\ul{\cal M}}$  in $d$-dimensional
spacetime with coordinates $Z^{\ul M}=(X^{\ul m}, \theta^{\ul \mu})$.
Following the
notation and conventions of [13], we shall always use underlined
indices for target superspace quantities. Let us define the pulled-back
supervielbein as
$$
E_\tau^{\ul A}=\partial_\tau Z^{\ul M} E_{\ul M}^{~\ul A}\ ,
\eqno(2.1) $$
where $\partial_\tau$ denotes differentiation with respect to the worldline
time
variable. The tangent space index splits as ${\ul A}=({\ul a},\ul{\alpha' r})$,
where ${\ul a}=0,1,...,d-1$ is the Lorentz vector index,
${\ul\alpha}=1,...,M$ labels the spinor irrep.~of the Lorentz group
and ${\ul r}=1,...,N$ labels the defining representation of the
automorpism group of the super-Poincar\'e algebra in $d$ dimensions.
For the sake of definiteness we shall consider the cases listed in Table 1.
In fact, they essentially correspond to the cases suggested by the
super $p$--brane  theories. Furthermore, to simplify the notation, we shall
denote the pairs of indices $\ul{\alpha' r}$ by a single index ${\ul \alpha}$,
e.g. $C_{\ul{\alpha\beta}}=C_{\ul{\alpha' r},\ul{\beta' s}}=
C_{\ul{\alpha'\beta'}}\eta_{\ul{rs}}$.

We next introduce the super
one form $B=dz^{\ul M} B_{\ul M}$ whose tangent space components are
defined with the help of the inverse supervielbein as follows:
$B_{\ul A}=E^{\ul M}_{\ul A}B_{\ul M}$.
The action for a massive superparticle, whose mass we shall set equal to one,
can then be written as
$$
    S=\int d\tau \bigg(\ft12 e^{-1}E_\tau^{\ul a}
E_\tau^{\ul a}  +\ft12  e + E_\tau^{\ul A}B_{\ul A}\bigg)\ ,
\eqno(2.2) $$
where $e$ is the einbein on the worldline. This action is invariant under
the following $\kappa$--symmetry transformations
$$
 \delta Z^{\ul M} E_{\ul M}^{\ul a} = 0\ ,\quad\quad
   \delta Z^{\ul M} E_{\ul M}^{\ul{\alpha}} =
   (1+\Gamma)^{\ul{\alpha\beta}}\ \kappa_{\ul\beta}\ , \quad\quad
    \delta e= S^{\ul \alpha}\kappa_{\ul \alpha}\ ,\eqno(2.3)
$$
where
$$
\Gamma^{\ul{\alpha\beta}} =-{1\over e}
\ E^{\ul a}_\tau(\Gamma_{\ul a})^{\ul{\alpha\beta}}\ , \quad\quad
S^{\ul \alpha}=4iE_\tau^{\ul \alpha}+2 E_\tau^{\ul a}
\big(u^{\ul \alpha}{}_{\ul a}
+\Gamma_{\ul a}^{\ul{\alpha\beta}}\ v_{\ul\beta}\big) \ .
\eqno(2.4)
$$
and $u^{\ul\alpha}{}_{\ul a}$ is an arbitrary  {\it
$\Gamma$--traceless}  vector-spinor superfield, $v_{\ul \alpha}$ is an
arbitrary  spinor superfield and $C_{\ul{\alpha\beta}}$ is the
charge conjugation matrix [19]. The invariance of the action imposes the
following torsion $T$ and $H=dB$ constraints
$$
\eqalign{
&T_{\ul{\alpha\beta}}{}^{\ul c}=
-2i(\Gamma^{\ul c})_{\ul{\alpha\beta}}\ ,\quad\quad
T_{{\ul\alpha} ({\ul {bc}})}
=u^{\ul \beta}{}_{({\ul b}}\Gamma_{{\ul c}){\ul{\beta\alpha}}}
+\eta_{\ul {bc}} v_{\ul\alpha}\ ,
\cr
&H_{\ul{\alpha\beta}}=-2i C_{\ul{\alpha\beta}}\ ,
\quad\quad H_{\ul {\alpha a}}=
(\Gamma_{\ul a})_{\ul{\alpha\beta}}\ v^{\ul\beta}
+ u_{\ul{\alpha a}}   \ .\cr}\eqno(2.5)
$$
 Notice that the
right hand side of the equation involving $H_{\ul{\alpha\beta}}$
must be symmetric. Therefore, $d=5,8,9$ are singled
out in Table 1. Lower values of $N$ can allow other dimensions (e.g.
$d=3$) which can be easily incorporated to the present scheme by ammending
our notation slightly. The physical interpretation of the
constraints (2.5)  requires a lengthy analysis of the Bianchi
identities, which we
hope to return to elsewhere. The expected
result is that they will be consistent with supergravity theories,
possibly coupled to a
matter/Maxwell sector in appropriate dimensions.

For future use, we also write down the Nambu-Goto form of the action, which
can be obtained from (2.2) by substituting the field equation of the einbein:
$$
S=\int d\tau\bigg[ (E_\tau^{\ul a}E_\tau^{\ul a})^{1/2}
+E_\tau^{\ul A}B_{\ul A}\bigg]\ . \eqno(2.6)
$$

Our purpose now is to reformulate the above theory in such a way that the
$\kappa$--symmetry is traded for worldline local supersymmetry. Since
the $\kappa$--symmetry parameter has $MN$ real components, and due to
the usual argument that only half of them count as true gauge
transformation parameters, it follows that the maximum world-line extended
symmetry to expect is $\ft12 MN$. From Table 1, we see that for $d=9,8,5$
we have $n=8$. Thus let us elevate
the world-line to a super worldline ${\cal M}$ with coordinates $Z^M=(\tau,
\theta^\mu),\ \mu=1,...8$. Following refs. [10][13], we shall take
${\cal M}$
to be superconformally flat. (We refer to ref. [10][13] for a detailed
geometrical description of such a space). In particular, the components of the
super torsion $T^C_{AB}$ will be those of a flat $n$--extended
world-line superspace:
$$
T_{rs}{}^0=-2i\delta_{rs}\ ,\quad\quad T_{0r}{}^0=0\ ,\quad\quad
T_{s0}{}^r=0\ ,\quad\quad T_{rs}{}^q=0\ .  \eqno(2.7)
$$
The super world-line tangent space index $A$ splits as $A=(0, r),\
r=1,...,8$.
As shown in [10][13], the superdiffeomorphisms which preserve these
constraints take the form
$$
\eqalign{
\delta\tau &=\lambda-\ft12 \theta^r D_r\lambda\ , \cr
\delta\theta_r &= -\ft{i}2 D_r\lambda\ , \cr}\eqno(2.8)
$$
where $\lambda$ is an arbitary superfield. These transformations contain
the world-line diffeomorhisms and the $n=8$ local world-line
supersymmetry. Under these transformations, the covariant derivative $D_r$
transforms homogeneously.

 The change of variable, which is
sometimes referred to as the twistor constraint, which is  needed to pass
from the $\kappa$--symmetric formulation to the world-line supersymmetric one,
is as follows
$$ \lambda_r^{\ul{\alpha}}\
\Gamma^{\ul a}_{\ul{\alpha\beta}}\ \lambda_s^{\ul{\beta}}
=\delta_{rs}E_0^{\ul a}\ , \eqno(2.9)
$$
where $\lambda_r^{\ul{\alpha}}$ are {\it commuting} spinors referred to as the
twistor variables and $E_0^{\ul a}=E_0^{~M}\partial_M Z^{\ul M}
E_{\ul M}^{~\ul a}$.
The strategy is to arrange that this equation arises as the $\theta_r=0$
component of an appropriate superfield equation. To this end, it is
convenient to define the matrix
$$
E_A^{~\ul A}=E_A^{~M}\big( \partial_M Z^{\ul M}\big) E_{\ul M}^{~\ul A}\ .
\eqno(2.10)
$$
Using this matrices, we can write the desired superfield equation as
$$
      \big(E_r\Gamma^{\ul a} E_s\big) =\delta_{rs} E_0^{\ul a}, \eqno(2.11)
$$
with the identifications
$$
   E_r^{\ul{\alpha}}\vert_{\theta=0}=\lambda_r^{\ul{\alpha}}\ ,\quad\quad
  E_0^{\ul a}\vert_{\theta=0}=E_0^{\ul a}\ . \eqno(2.12)
$$
We use a notation in which the contracted $\ul{\alpha}$ indices are
suppressed, and the paranthesis in (2.11) indicate that such contractions are
made. In flat superspace, these identifications mean that
$\theta^{\ul \alpha}(\tau,\theta)=\theta^{\ul\alpha}(\tau)
+\lambda_r^{\ul\alpha}(\tau)\theta^r+\cdots $, i.e. the twistor variable
$\lambda_r^{\ul\alpha}$ is the superpartner of the target superspace fermionic
coordinate $\theta^{\ul\alpha}(\theta)$.

{}From the identity (2.11) it follows that
$$
      (E_r\Gamma^{\ul a} E_s) =\ft18 \delta_{rs}
      (E_q\Gamma^{\ul a} E_q) \ .\eqno(2.13)
$$
This identity has arisen in the twistor formulation of
string theory in $d=10$, and its group theoretic interpretation has been
given in [8]. Its dimensional reduction from $d=10$ down to
$d=9,8,5$ yields, in addition to the corresponding twistor
identities  of the form (2.13), other identities as well. In particular,
the following identity will arise
$$
(E_r E_s)=\ft18 \delta_{rs} (E_qE_q)  \ . \eqno(2.14)
$$

Our task is to
write an action in $n=8$  world-line superspace which will

a) give rise to the constraints (2.11) and (2.14),

b) given the constraints (2.5) and (2.7), the action will possesss
world-line $n=8$ local supersymmetry.

To this end, we propose the following action
which is the appropriate generalization for a massive
superparticle of the action
given for the massless superparticle in refs.~[10][13]:
$$
S= \int d\tau d^8 \theta\bigg[ P_{\ul a}^r E_r^{\ul a}
+P^M({\tilde B}_M -\partial_M Q)\bigg]\ ,  \eqno(2.15)
$$
where $P_{\ul a}^r$, $P^M$ and $Q$ are Lagrange multiplier superfields and
${\tilde B}_M$ is defined by
$$
{\tilde B}_M= \partial_M Z^{\ul M}B_{\ul M} -{i\over 16}E_M^0 H_{rr}
\ , \eqno(2.16)
$$
where $H_{rr}=E_r^{\ul A}E_r^{\ul B} H_{\ul{BA}}$
and $H_{\ul{BA}}$ are the tangent space components of the field strength
$H=dB$:
$$H_{\ul {AB}}= (-)^{\ul {A}(\ul {B} + \ul {N})} E_{\ul {B}}^{\ul {N}}
E_{\ul {A}}^{\ul {M}} H_{\ul {MN}}\ , \eqno(2.17)
$$
where the indices in the exponent indicate Grassmannian parities. Recall
that $M=(\tau,\mu)$, $A=(0,r)$, ${\ul M}=(\ul m,\ul \mu)$ and ${\ul A}=(\ul
a,\ul \alpha)$. The indices of the bosonic (fermionic) coordinates have the
parity 0(1). This is consistent with the fact that the twistor variables
$\lambda_r^{\ul \alpha}$ are commuting variables.

	The above form of ${\tilde
B}$ is engineered such that $d{\tilde B}=0$ modulo the constraints
(2.5),(2.7),(2.11) and (2.14), as we shall show below. Note that, the
independent world-line {\it superfields} are: $P_{\ul a}^r,\ P^M,\ Q,\
E_M^{~A}$ and $Z^{\ul M} $.  An important property of the action (2.15) is
that it is invariant under $n=8$ local  world-line supersymmetry, as opposed
to the $\kappa$--symmetry. (The latter emerges as a special case of the former
in a certain gauge). The supersymmetry of the second and third terms in the
action is manifest (everything transform like supertensors), while the
supersymmetry of the first term is due to the fact that $E_r^{\ul a}$
transforms homogeneously like $D_r$ does, and this can be compensated by a
suitable transformation of the Lagrange multiplier.

 At this stage, to simplify matters, we
shall set the inconsequential superfields $u^{\ul\alpha}{}_{\ul a}$
and $v_{\ul \alpha}$  in (2.5)
equal to zero, and take the resulting  constraints and their
Bianchi consequences to characterize the target space background.
Thus we have the constraints,
$$
\eqalign{
&T_{\ul{\alpha\beta}}{}^{{\ul c}}=
-2i(\Gamma^{\ul c})_{\ul{\alpha\beta}}\ ,\quad\quad
T_{\ul{b\alpha}}{}^{\ul a}=0\ , \quad\quad
T_{\ul{\alpha\beta}}{}^{\ul\gamma}=0\ , \cr
&H_{\ul{\alpha\beta}}=-2i C_{\ul{\alpha\beta}}\ ,
\quad\quad H_{\ul{a\alpha}}=0\ . \cr} \eqno(2.18)
$$

With eqs.~(2.7) and (2.18) at hand, we can now analyze the content of the
superfield equations that follow from the action (2.15). Firstly, the
equation of motion for $P_{\ul a}^r$ is simply
$$
E_r^{\ul a}=0\ .  \eqno(2.19)
$$
The supercovariant derivative of this equation in the spinorial direction
evaluated at $\theta^r=0$ gives the desired constraint (2.11). To see this,
it useful first to evaluate the curl of $E_A^{\ul A}$ defined in (2.10). We
find:
$$
D_A E_B^{~\ul C}-(-1)^{AB}D_B E_A^{~\ul C}=-T_{AB}{}^C
E_C^{\ul C}+(-1)^{A(B+{\ul D})} E_B^{~\ul D}
E_A^{~\ul E}T_{{\ul E}{\ul D}}{}^{\ul C} \ , \eqno(2.20)
$$
where the covariant derivative $D_A=E_A^{~M}D_M$ rotates the indices $A$
and ${\ul A}$ and the tangent space components of the supertorsion
$T_{MN}{}^C=\partial_M E_N^{~C}+\Omega_M^{CD}
E_{ND}-(-1)^{MN}(M\lra N)$ are defined as follows
$$
T_{AB}{}^C=(-1)^{A(B+N)}E_B^{~N}E_A^{~M}T_{MN}{}^C \ , \eqno(2.21)
$$
and similarly for $T_{{\ul A}{\ul B}}{}^{\ul C}$.
Taking the spinor-spinor component of (2.20) and using the
constraints (2.7), (2.14) and (2.19) we indeed obtain the twistor
constraint
equation (2.11). The $\theta_r=0$ component of the equation gives (2.9) and
one can show that there is no further information coming from the higher
order $\theta$ expansion.

We next consider the equation of motion for the Lagrange multiplier $P^M$
which simply reads
$$
    H_{MN} = \partial_M{\tilde B}_N-(-1)^{MN}\partial_N{\tilde B}_M=0\ .
\eqno(2.22)
$$
This equation, together with (2.19), is at the center of the
construction of the model. Defining ${\tilde H}=d{\tilde B}$,
and referring to its tangent space components, we obtain
$$
\eqalign{
{\tilde H}_{AB}=&(-1)^{A(B+{\ul B})}E_B^{~\ul B}E_A^{~\ul A}H_{\ul{AB}}
 -{i\over 16} (-1)^{A(B+N)}E_B^{~N}E_A^{~M}\ \times\cr
&\bigg [ \partial_M\big(E_N^{~0}H_{rr}\big) - (-)^{MN}
        \partial_N\big(E_M^{~0}H_{rr}\big)  \bigg ]
=0\ . \cr} \eqno(2.23)
$$
where the (world-sheet) tangent space components $H_{AB}$ are related to the
(target space) tangent space components $H_{\ul {AB}}$ according to
$$
H_{AB}=(-1)^{A(B+{\ul B})}E_B^{~\ul B}E_A^{~\ul A}H_{\ul{AB}}\ . \eqno(2.24)
$$
We can write (2.23) as
$$
{\tilde H}_{AB}=H_{AB}-{i\over 16}T_{AB}{}^0H_{rr}
+{i\over 16}\bigg[ \delta_A^0 D_B
H_{rr}-(-1)^{AB}\delta_B^0 D_A H_{rr}\bigg] = 0 \ . \eqno(2.25)
$$
Taking the spinor-spinor component of this equation gives
$$
H_{rs}-\ft18 \delta_{rs} H_{qq}=0\ .  \eqno (2.26)
$$
{}From (2.24) and (2.18), we see that $H_{rs}=-2i(E_r E_s)$, and hence
(2.14) follows from (2.26). Thus, we shall
consider (2.14) to follow from the integrability condition of the $P^M$
equation of motion.

Next, we consider the time-spinor projection of (2.25). It yields,
$$
H_{0r}+{i\over 16}D_r H_{qq}=0\ . \eqno(2.27)
$$
This equation is precisely what one obtains by considering the Bianchi
identity $D_{(r}H_{st)} - T^0_{(rs}H_{t)0}=0$ and using equations
(2.7) and (2.26). Therefore, (2.27) is satisfied as well without implying new
constraints. This concludes the proof that indeed $d{\tilde B}=0$. As a
consequence of this property, the action (2.15) has also the gauge
invariance
$$
\delta P^M=\partial_N \Lambda^{NM}\ , \eqno(2.28)
$$
where $\Lambda^{MN}$ is an arbitrary graded antisymmetric superfield. In
showing this invariance we need to use (2.22), which in turn involves the
use of the constraint (2.19). This
constraint follows as the field equation of the Lagrange multiplier
$P_r^{\ul a}$. Such terms can be cancelled by
an appropriate variation of the Lagrange multiplier $P_r^{\ul a}$. Therefore,
(2.28) is  indeed a symmetry of the action.

Next, we consider the equation of motion for $Q$ which reads
$$
\partial_M P^M=0 \ . \eqno(2.29)
$$
This equation has the solution [13]
$$
P^M=\partial_N\Sigma^{NM}+\theta^8\delta_\tau^M T\ , \eqno(2.30)
$$
where $T$ is a constant and $\Sigma^{MN}$ is an arbitrary graded
antisymmetric superfield. Substituting $P^M=\theta^8 \delta^M_\tau T$ into the
action (2.15) yields (setting $T=1$ )
$$
\eqalign{
S= & \int d\tau d^8\theta P_{\ul a}^r E_r^{\ul a}\ +\int d\tau
{\tilde B}_\tau\vert_{\theta=0}  \cr
=&\int d\tau d^8\theta P_{\ul a}^r E_r^{\ul a}\
+\int d\tau\bigg[ \partial_\tau Z^{\ul M}B_{\ul M}
-{1\over 8}E_\tau^0 (\lambda_r\lambda_r)\bigg] \ . \cr}\eqno(2.31)
$$
To simplify this further, consider the following Dirac matrix
identity which holds in $d=5,9$
$$
\Gamma^{\ul a}_{\ul{\alpha\beta}}\Gamma^{\ul a}_{\ul{\gamma\delta}}
+C_{\ul{\alpha\beta}}C_{\ul{\gamma\delta}}
+{\rm cyclic}\ (\ul{\alpha\beta\gamma}) =0\ . \eqno(2.32)
$$
Multiplying this equation by $E_r^{\ul\alpha}E_r^{\ul\beta}E_s^{\ul\gamma}
E_s^{\ul\delta}$ and using the identities (2.13) and (2.14), we find that
$(E_rE_r)=-8\big(E_0^{\ul a}E_0^{\ul a}\big)^{1/2}$. Evaluating this at
$\theta=0$ and substituting the result into (2.31), we find
$$
S=\int d\tau d^8\theta P_{\ul a}^r E_r^{\ul a}\
+\int d\tau\bigg[\partial_\tau Z^{\ul M}B_{\ul M}
+E_\tau^0 \big(E_0^{\ul a}
E_0^{\ul a}\big)^{1/2}\bigg]\ . \eqno(2.33)
$$
Note that
$$
E_\tau^0 E_0^{\ul a}=
E_\tau^A E_A^{\ul a}=E_\tau^{\ul a}\ , \eqno(2.34)
$$
modulo the constraint (2.19). The effect of using the constraint
(2.19) in the action amounts to a redefinition of
the Lagrangian multiplier $P_{\ul a}^r$. Therefore, using (2.34) we can
simplify the last term in the action and obtain
$$
S=\int d\tau d^8\theta P_{\ul a}^r E_r^{\ul a}\
+\int d\tau\bigg[ \partial_\tau Z^{\ul M}B_{\ul M}
+\big(E_\tau^{\ul a}E_\tau^{\ul a}\big)^{1/2}\bigg] \ .
\eqno(2.35)
$$
We see now that the second integral in (2.35) agrees  with the
$\kappa$--symmetric action (2.6).  Finally, following the same
arguments as in [13], the component form of the first term in the action can
also be computed and one finds the following component action
$$
S=\int d\tau\bigg[ p_{\ul a}\big(E_0^{\ul a}
-\ft18 (\lambda_r\Gamma^{\ul a} \lambda_r) \big)
 +\partial_\tau Z^{\ul M}B_{\ul M}
+\big(E_\tau^{\ul a}E_\tau^{\ul a}\big)^{1/2}\bigg] \ ,
\eqno(2.36)
$$
where $p_{\ul a}=(D^7)_r P_{\ul a}^r\vert_{\theta=0}$. With arguments
parallel to those of [10][13], we expect that the  Lagrange multiplier
$p_{\ul a}$ does not describe any new degree of freedom, and the field
equations of (2.6) and (2.36) are classically equivalent. In the case of
the massless superparticle, showing this equivalence requires the use of an
important Abelian gauge symmetry [10]. A generalized version of this
symmetry is also present in the massive superparticle case. We find that the
action (2.15) is invariant under the gauge transformations
$$
\delta P_{\ul a}^r = D_q\bigl(\xi^{qrs}\Gamma_{\ul a} E_s\big)\ ,\quad\quad
\delta P^M = -E_r{}^M D_q\bigl(\xi^{qrs} E_s\big)\ , \eqno(2.37)
$$
where the parameter $\xi^{qrs}_{\ul\alpha} (\tau,\theta)$ is totally
symmetric and traceless in its worldline indices. Note that, unlike in the
massless particle case, both of the lagrange multipliers transform here. To
show that this is an invariance of the action, we need to use the Dirac matrix
identity (2.33) and the constraints (2.7) and (2.18), which imply the target
space equations of motion.

The rest of the paper will be devoted to a discussion of the twistor-like
formulation of super $p$--branes with $p\ge 2$.
\bigskip
\noindent{\bf 3. Super $p$--Branes ($p\ge 2$) }
\bigskip
The $\kappa$--symmetric formulation of super $p$--branes is well known
[20-22].
Here, we shall directly investigate the construction of a worldvolume
locally supersymmetric version along the lines of the massive
superparticle case described in detail above. The relevant target spaces
are listed in Table 1. As before, the maximum number of real
components of the worldvolume supersymmetry parameter is $\ft12 MN$. This
translates into $n=2,4,8$ supersymmetry in various cases as indicated in
Table 1.

The coordinates of the worldvolume superspace ${\cal M}$ are $Z^M=(X^m,
\theta^\mu),\ m=1,...,p+1,\ \mu=1,...,\ft12 MN$. The supervielbein is
again denoted by $E_M^{~A}$ with the tangent space indices splitting as
$A=(a,\alpha' r),\ a=1,...,p+1,\ \alpha'=1,...,m,\ r=1,...,n$. (See Table 1).
For simplicity in notation, {\it we\ will\ indicate\ the\ pair\ of\ indices
$\alpha' r$\ by\ a\ single\ index\ $\alpha$.} Following [16], we shall take
${\cal M}$
to be characterized by the following super torsion constraints
$$
T_{\alpha\beta}{}^a=-2i (\Gamma^a)_{\alpha\beta},\qquad
T_{b\alpha}{}^a=0,\qquad T_{bc}{}^a=0,\qquad
T_{\alpha\beta}{}^{\gamma}=0\ . \eqno(3.1)
$$
See Table 1 for the symmetry properties of the gamma matrices. In
particular note that, $\Gamma_{\alpha\beta}^a=
\Gamma_{\alpha'\beta'}^a\eta_{rs}$ where $\eta_{rs}$ is the
invariant tensor of the  automorphism group $G$. Thus,
$\eta_{rs}$ is the unit matrix $\delta_{rs}$ when $G$ is an orthogonal group,
and the constant antisymmetric matrix $\Omega_{rs}$ when $G$ is a symplectic
group. From Table 1 we see that $\eta_{rs}=-\epsilon \eta_{sr}$ and
$\Gamma_{\alpha'\beta'}=-\epsilon \Gamma_{\beta'\alpha'}$, with $\epsilon=-1$
for
orthogonal $G$ and $\epsilon=1$ for symplectic $G$. Similar properties hold for
the corresponding target space quantities.

The coordinates of the target superspace $\ul{\cal M}$ are
 $Z^{\ul M}=(X^{\ul m},\theta^{\ul \mu}),\ {\ul m}=0,...,d-1,\
{\ul\mu=1,...,MN}$ (See Table 1). The supervielbein is
 $E_{\ul M}^{~\ul A}$ with the tangent space index splitting as
${\ul A}=({\ul a},{\ul \alpha}),\ {\ul a}=0,...,d-1,\ \ul{\alpha}=1,...,MN$.
The index ${\ul\alpha}$ is short for a pair of indices $(\ul{\alpha' r})$,
with ${\ul\alpha'}=1,....M,\ {\ul r}=1,...,N$. The superspace
$\ul {\cal M}$ is
also endowed with a super $(p+1)$--form $B$ whose curvature is $H=dB$.

The $\kappa$--symmetry of the usual super $p$--brane action imposes
constraints
on the torsion and the $(p+2)$--form $H$ [21]. As before, arbitrary
superfields $u^{\ul\alpha}{}_{\ul a}$ and $v_{\ul \alpha}$ arise [21][19].
As we
did in the particle case, we shall set these inconsequential superfields equal
to zero, and furthermore we shall fix the target space supergeometry, in a
manner which is consistent with $\kappa$--symmetry, to be characterized by the
following constraints
$$
\eqalign{
&T_{\ul{\alpha\beta}}{}^{\ul c}=
-2i(\Gamma^{\ul c})_{\ul{\alpha\beta}}\ ,\quad\quad
T_{\ul{b\alpha}}{}^{\ul a}=0\ , \quad\quad
T_{\ul{\alpha\beta}}{}^{\ul\gamma}=0\ , \cr
&H_{\ul{\alpha\beta c_1}...\ul{c_p}}=
i\xi^{-1}\big(\eta\Gamma_{\ul{c_1}...\ul{c_p}}\big)_{\ul{\alpha\beta}},\qquad
H_{\ul{\alpha b_1}...\ul{b_{p+1}}}=0,\qquad
H_{\ul{\alpha\beta\gamma}...\ul{A_1}...{\ul{A_{p-1}}}}=0\ , \cr} \eqno(3.2)
$$
where $\xi=(-)^{(p-2)(p-5)/4}$ and $\eta$ is a matrix chosen such that
$\eta\Gamma_{\ul{c_1}...\ul{c_p}}$ is symmetric. $\eta=1$ except for the
following cases: $\eta=\Gamma_{d+1}$ for $(p=3, d=8)$, with the definition
$\Gamma_{d+1}=\Gamma_0\Gamma_1\cdots \Gamma_{d-1}$, and $\eta=1\times
\sigma_2$ for $(p=2, d=5)$. See the Table for further information on the
notation and properties
   of the Dirac matrices in
diverse dimensions.

In $d=11$ dimensions the above constraints
describe the $d=11$ supergravity theories.
In other cases, a detailed
analysis of the constraint remains to be carried out. Presumably, they
describe supergravity theories containing $(p+1)$--form potentials.

Having specified the geometry of the worldvolume and target superspaces,
our next goal is to write down an action for
twistor--like super $p$--branes in analogy with the action (2.15). Such an
action has already been proposed in [16] for the case of the supermembrane.
Here
we generalize that result and propose the following action for all super
$p$--branes:
$$
S= \int d^{p+1}\sigma d^{mn}\theta\bigg[ P_{\ul a}^\alpha
E_\alpha^{\ul a}  +P^{M_1\cdots M_{p+1}}\big({\tilde B}_{M_1\cdots M_{p+1}}
-\partial_{M_1} Q_{M_2\cdots M_{p+1}}\big)\bigg]\ ,  \eqno(3.3)
$$
where $P_{\ul a}^{\alpha r}$, $P^{M_1\cdots M_{p+1}}$ and
$Q_{M_1\cdots M_p}$ are Lagrange multiplier superfields (the latter two
are graded totally antisymmetric) and the $(p+1)$--form ${\tilde B}$ is
given by
$$
\eqalign{
{\tilde B}_{M_1\cdots M_{p+1}}=&(-1)^{\epsilon_{p+1}(M,\ul{M})}\
 \partial_{M_{p+1}}Z^{\ul{M_{p+1}}}\cdots
 \partial_{M_1}Z^{\ul M_1} B_{\ul{M_1}\cdots \ul{M_{p+1}}}\cr
  &-{i\over 2mn(p+1)}\Gamma^{\alpha\beta}_{c_{p+1}}
          \bigg(E_{M_{p+1}}^{~c_{p+1}}\cdots E_{M_1}^{~c_1}
H_{\alpha\beta c_1\cdots c_p}+ {\rm cyclic}
\ [M_1\cdots M_{p+1}] \bigg)\ .\cr} \eqno(3.4)
$$
The grading factor is given by
$$
\epsilon_{p+1}(M,\ul{M})=
\sum_{n=1}^{p}(M_1+\cdots M_n)(M_{n+1}+\ul{M_{n+1}})\ , \eqno(3.5)
$$
and the pullback of $H$ by
$$
H_{A_1\cdots A_{p+2}}=(-1)^{\epsilon_{p+2}(A,\ul{A})}
E_{A_{p+2}}^{~\ul{A_{p+2}}}\cdots E_{A_1}^{~\ul{A_1}}
H_{\ul{A_1}\cdots \ul{A_{p+2}}}\ . \eqno(3.6)
$$

The field equation for $P_{\ul a}^\alpha $ is
$$
    E^{\ul a}_\alpha =0\ . \eqno(3.7)
$$
The integrability condition for this equation yields the analog of the
twistor constraint (2.13) for super $p$--branes. It can be obtained from
(2.20), (3.1), (3.2) and takes the form
$$
(E_\alpha\Gamma^{\ul a} E_\beta )=\Gamma^a_{\alpha\beta} E_a^{\ul a}\ .
\eqno(3.8)
$$
Recall that $\Gamma_{\alpha\beta}^a=
\Gamma_{\alpha'\beta'}^a\eta_{rs}$. We shall use (3.7) and (3.8) repeatedly in
the following calculations.

The field equation for $P^{M_1\cdots M_{p+1}}$ is
$$
{\tilde H}_{M_1 \cdots M_{p+2}} =
\partial_{M_1}{\tilde B}_{M_2\cdots M_{p+2}} +{\rm cyclic}\ [M_1\cdots
M_{p+2}]\ =0 \ .  \eqno(3.10)
$$
Given ${\tilde B}$ as in (3.4), it is nontrivial to show
(3.10). To this end, we first refer to the tangent space components of
(3.10)
which read
$$
\eqalign{
{\tilde H}_{A_1 \cdots A_{p+2}} =&\ H_{A_1\cdots A_{p+2}} - \bigg (
{i\over 2mn}\Gamma_{c_1}^{\alpha\beta}
\ T_{A_1A_2}{}^{[c_1}\ \delta^{c_2}_{A_3}\cdots \delta_{A_{p+2}}^{c_{p+1}]}
\ H_{\alpha\beta c_2\cdots c_{p+1}} \cr
+&{i(-1)^{p+1}\over 2mn(p+1)}\ \Gamma^{\alpha\beta}_{c_1}
\delta^{c_1}_{A_1}\cdots \delta_{A_{p+1}}^{c_{p+1}}\ D_{A_{p+2}]}
H_{\alpha\beta c_2\cdots c_{p+1}}\cr
+&{\rm cyclic}\ [A_1\cdots A_{p+2}]  \bigg ) = 0\ .\cr}\eqno(3.11)
$$
Using (3.1), (3.2) and (3.7) we find that all the projections of
${\tilde H}$ are identically vanishing except
${\tilde  H}_{\alpha\beta c_1\cdots c_p}$
and ${\tilde H}_{\alpha c_1\cdots c_{p+1}}$. The vanishing of the former
gives the equation
$$
H_{\alpha\beta c_1\cdots c_p}
={1\over mn(p+1)}\Gamma_{\alpha\beta}^{c_{p+1}}\bigg[
\Gamma_{c_{p+1}}^{\gamma\delta}H_{\gamma\delta c_1\cdots c_p}
+{\rm cyclic}\ [c_1\cdots c_{p+1}]\bigg]\ . \eqno(3.12)
$$
We observe that the expression in the square brackets is totally antisymmetric
in  $(c_1\cdots c_{p+1})$ and therefore
it must be proportional to the Levi-Civita symbol
$\epsilon^{c_1\cdots c_{p+1}}$. Thus we can write
$$
H_{\alpha\beta c_1\cdots c_p}
=\epsilon_{c_1\cdots c_{p+1}}\Gamma^{c_{p+1}}_{\alpha\beta} Q\ ,\eqno(3.13)
$$
for some $Q$. Introducing the notation
$$
H_{\alpha\beta c_1\cdots c_p} :=
\epsilon_{c_1\cdots c_p a} H^a_{\alpha\beta}\ ,
\eqno(3.14)
$$
we can write (3.13) as
$$
    H_{\alpha\beta}^a=\Gamma_{\alpha\beta}^a Q\ . \eqno(3.15)
$$

{}From the definition of the pull-back of $H$ and using the
constraints (3.2) we have
$$
H_{\alpha\beta c_1\cdots c_p}=i\xi^{-1}
E_{c_1}^{~\ul c_1}\cdots E_{c_p}^{~\ul c_p}
\big(E_\alpha\eta\Gamma_{\ul{c_1}\cdots \ul{c_p}} E_\beta\big)\ .
\eqno(3.16)
$$
Using this equation, we now have to show that (3.15) is satisfied. Our
strategy is to replace one of the $E_a^{~\ul c}$ factors in (3.16), by using
the following identity which follows from (3.8)
$$
E_a{}^{\ul a}={1\over mn}\Gamma_a^{\alpha\beta}\big(E_\alpha
\Gamma^{\ul a} E_\beta \big)\ , \eqno(3.17)
$$
and then making use of the super $p$--brane Dirac matric identity
$$
\Gamma^{\ul c}_{(\ul{\alpha\beta}} \big(\eta \Gamma^{\ul{c c_1}\cdots
\ul{c_{p-1}}}\big)_{\ul{\gamma\delta})}=0\ . \eqno(3.18)
$$
 In this fashion,  after a little bit of algebra, from (3.16) we obtain
$$
\eqalign{
 p\big(mn+4\big)H^a_{\alpha\beta}=&
 \bigg(  H^c_{\gamma\delta}
\ \Gamma_c^{\gamma\delta}\ \Gamma^a_{\alpha\beta}-
H^a_{\gamma\delta}
\ \Gamma_c^{\gamma\delta}\ \Gamma^c_{\alpha\beta}      \bigg)\cr
 &+2H^c_{\alpha\gamma}(\Gamma_c{}^a)^\gamma{}_\beta
  +2H^c_{\beta\gamma}(\Gamma_c{}^a)^\gamma{}_\alpha\ . \cr}\eqno(3.19)
$$
We now decompose $H^a_{\alpha\beta}$ as follows:
$$
H^a_{\alpha\beta}=\Gamma^a_{\alpha\beta}Q
+ H^{ab}(\Gamma_b)_{\alpha\beta} + {\hat H}^a_{\alpha\beta} \eqno (3.20)
$$
with $H^{ab}$ traceless in $ab$ and
$$
(\Gamma^a)^{\alpha\beta}{\hat H}^b_{\alpha\beta} = 0\ .\eqno(3.21)
$$
Substituting the above parametrization of $H^a_{\alpha\beta}$
into (3.19), after some calculation, we find that $Q$ is not determined,
and that the expansion coefficient $H^{ab}$ is equal to zero.
This leaves us with the following equation for
${\hat H}^a_{\alpha\beta}$:
$$
p(mn+4) {\hat H}^a_{\alpha\beta} =
 2{\hat H}^c_{\alpha\gamma}(\Gamma_c{}^a)^\gamma{}_\beta
  +2{\hat H}^c_{\beta\gamma}(\Gamma_c{}^a)^\gamma{}_\alpha\ .
\eqno(3.22)
$$
We now rewrite the 2-gamma matrix in the second term on the r.h.s.
of (3.22) as $(\Gamma_c{}^a)^\gamma{}_\alpha =
(\Gamma_c{}^a)_\alpha{}^\gamma$ (see Table 1 for the symmetry of gamma
matrices). Next we write the 2-gamma matrices in eq.~(3.22) as products of
1-gamma matrices and multiplying this equation with $\Gamma_a^{\beta\eta}$ we
obtain
$$
\big [ p(mn+2)+2 \big ] \big ( {\hat H}^a_{\alpha\beta}
\Gamma^{\beta\eta}_a\big ) = 2 \big ({\hat H}^c_{\beta\gamma}
(\Gamma_c)_{\alpha\delta} \big )\big ((\Gamma^a)^{\delta\gamma}
\Gamma_a^{\beta\eta}\big ) \eqno(3.23)
$$
Contracting this equation with $\delta^\alpha_\eta$ we find
$$
{\hat H}^a_{\alpha\beta}\Gamma_a^{\beta\alpha} = 0 \ .
\eqno(3.24)
$$
It is convenient now to distinguish between
different values of $(p,m,n)$ (see Table).
We first consider the cases with $p=2$, i.e.~$(p,m,n)=
(2,2,8), (2,2,4), (2,2,2)$ or $(2,2,1)$. At this point it is useful
to write out explicitly
$\Gamma^a_{\alpha\beta}=\Gamma^a_{\alpha'\beta'}\delta_{rs}\ (\alpha'=1,2,\
r=1,...,n)$ where $\Gamma^a_{\alpha'\beta'}$ are the two by two Pauli
matrices. Multiplying (3.23) with $\delta_{\eta'}^{\alpha'}$ we find a stronger
version of (3.24) to hold namely:
${\hat H}^a_{\alpha' r,\beta' s}\Gamma_a^{\beta'\alpha'} = 0$. Using this
equation and using the fact that the Pauli matrices satisfy the relation
$$
\Gamma^a_{\alpha'\beta'}(\Gamma_a)_{\gamma'\delta'} =  \big (
C_{\beta'\gamma'}C_{\alpha'\delta'} + C_{\alpha'\gamma'}
C_{\beta'\delta'} \big )\ , \eqno(3.25)
$$
it is not too difficult to show that ${\hat H}^a_{\alpha\beta}=0$.
{}From the decomposition (3.20) then immediately follows the desired equation
(3.15).

We next consider the cases $(p,m,n)=(5,4,2)$ and $(3,4,1)$  where
the following gamma matrix identity holds:
$$
(\Gamma^a)^{\delta(\gamma} \Gamma_a^{\beta\eta)} = 0\ . \eqno(3.26)
$$
This identity is related to the construction of superstrings in
$d=6$ and $d=4$ target-space dimensions, respectively.
Using this identity, the fact that ${\hat H}_{\beta\gamma}=
{\hat H}_{\gamma\beta}$ and eq.~(3.21), it is then not too difficult to show
that again ${\hat H}^a_{\alpha\beta}= 0$, and thus (3.15) is indeed satisfied.

This leaves us with the cases $(p,m,n)=(4,4,2)$ and $(3,4,2)$. For these
case we can neither find a Fierz identity of the form (3.25) nor
can we in the (4,4,2) case apply the gamma-matrix identity (3.26) since
there are no superstrings in $d=5$ target-space dimensions. So far, we have
not been able to proof the identity (3.15) for these cases by other means.
This completes our discussion of the identity (3.15).

To complete the proof of (3.11) there remains to be shown that
${\tilde H}_{\alpha c_1\cdots c_{p+1}}$ vanishes. From (3.11) we obtain
$$
H_{\gamma c_1\cdots
c_{p+1}}={i\over 2mn(p+1)} \Gamma_{c_1}^{\alpha\beta}D_\gamma
H_{\alpha\beta s c_2\cdots c_{p+1}}+\ {\rm cyclic}\ [c_1\cdots
c_{p+1}]\ . \eqno(3.28)
$$
Introducing the notation
$$
 H_{\alpha c_1\cdots c_{p+1}} :=  \epsilon_{c_1\cdots c_{p+1}}Q_\alpha
\ , \eqno(3.29)
$$
and using (3.12), we can write (3.28) as follows
$$
    Q_\alpha ={i\over 2}(-1)^{p+1} D_\alpha Q\ . \eqno(3.30)
$$
To prove this equation, we consider the Bianchi identity
$D_{(\alpha} H_{\beta\gamma)c_1\cdots c_p}
+2i\Gamma^c_{(\alpha\beta}H_{\gamma)cc_1\cdots c_p}=0$. Using the
notations (3.13) and (3.29), this can be written as
$$
D_{(\alpha}H^a_{\beta\gamma)}
+2i(-1)^{p+1}\Gamma^a_{(\alpha\beta}Q_{\gamma)}=0. \eqno(3.31)
$$
Substituting (3.15) into this equation, we obtain the
equation we wanted to prove, namely (3.30). This completes the proof of
eq.~(3.11). As a consequence, the action (3.3) has the
additional symmetry
$$
     \delta P^{M_1\cdots M_{p+1}}=\partial_N \Sigma^{NM_1\cdots M_{p+1}}\ ,
\eqno(3.32)
$$
where the parameter is completely graded antisymmetric.

Now we turn to the equation of motion for the Lagrange multiplier
$Q^{M_1\cdots M_p}$ given by
$$
\partial_{M_1} P^{M_1\cdots M_{p+1}}=0\ . \eqno(3.33)
$$
In analogy with (2.31), using the gauge invariance (3.32), the solution of
the above equation can be put into the form
$$
P^{M_1\cdots M_{p+1}}=T\epsilon^{m_1\cdots m_{p+1}}
\delta_{m_1}^{M_1}\cdots \delta_{m_{p+1}}^{M_{p+1}} \theta^{mn}\ .\eqno(3.34)
$$
Substituting this into the action (3.3), we obtain (with $T=1$)
$$
\eqalign{
S=& \int d^{p+1}\sigma d^{mn}\theta P_{\ul a}^\alpha
E_\alpha^{\ul a}  +{i\over 2}(p+1)!\int d^{p+1}\sigma \big({\rm det}\
E_m^a\big) Q\vert_{\theta=0} \cr
 &+\int d^{p+1}\sigma \epsilon^{m_1\cdots m_{p+1}}
\partial_{m_{p+1}}Z^{\ul{M_{p+1}}}\cdots
 \partial_{m_1}Z^{\ul M_1} B_{\ul{M_1}\cdots
\ul{M_{p+1}}}\vert_{\theta=0} \cr}\ .
\eqno(3.35)
$$

The last term coincides with the Wess-Zumino term of the usual super
$p$--brane action, but to show that the second term is the Nambu-Goto
term requires quite a bit of further work. To this end, it is convenient to
introduce the following matrix
$$
\Gamma={\xi\over \sqrt{ -g}(p+1)!}\epsilon^{c_1\cdots c_{p+1}}
E_{c_1}^{~\ul{a_1}}\cdots E_{c_{p+1}}^{~\ul{a_{p+1}}}
\Gamma_{\ul{a_1}\cdots \ul{a_{p+1}}}\ , \eqno(3.37)
$$
where $g={\rm det}\ g_{ab}$, with the definition
$$
g_{ab}=E_a^{~\ul c}E_b^{~\ul c}\ .  \eqno(3.38)
$$
Using (3.38) one can easily show that $\Gamma^2=1$. Next, we define the
matrix
$$
\tau_a =E_a{}^{\ul a}\Gamma_{\ul a}\ ,  \eqno(3.39)
$$
which satisfies
$$
\{\tau_a,\tau_b\}=2g_{ab}\ ,
\qquad [\tau_a,\Gamma]=0\ ,\qquad\qquad
\tau_{c_1\cdots c_p}=-\xi^{-1}\epsilon_{c_1\cdots c_{p+1}}
\Gamma\tau^{c_{p+1}}\sqrt{ -g}\ .\eqno(3.40)
$$
Using (3.39) and (3.16) we can write (3.13) as
$$
(E_\alpha\eta\tau_{c_1\cdots c_p}E_\beta)=
-i\xi\epsilon_{c_1\cdots c_p a}\Gamma^a_{\alpha\beta} Q \ .\eqno(3.41)
$$
We now derive an identity for $Q$. Multiplying
the cyclic identity (3.18) by
$\epsilon^{c_1\cdots c_{p-1}ab}\Gamma_a^{\alpha\beta}$\hfill\break
$\Gamma_b^{\gamma\delta}
E_{c_1}^{~\ul c_1}\cdots E_{c_{p-1}}^{~\ul{c_{p-1}}}
\bigl(E_\alpha \Gamma\eta\bigr)^{\ul \alpha}
E_\beta^{\ul\beta}E_\gamma^{\ul\gamma} E_\delta^{\ul\delta}$, and then using
the equations (3.8), (3.37), (3.39), (3.40) and (3.41)  we find that
$$
Q^2= {\rm det}\ g
+{2(-)^{(p+1)}\xi\sqrt {-g}\epsilon^{c_1\cdots c_{p-1}ab}\over
(mn)^2(p+1)!} \big ( E_\alpha\eta\Gamma\Gamma^{\ul {a}} E_\delta)
\Gamma_a^{\alpha\beta}\Gamma_b^{\gamma\delta}
 \big (E_\gamma\eta\Gamma^{\ul a}{}_{c_1\cdots c_{p-1}}E_\beta \big)\ ,
\eqno(3.42)
$$
where $\Gamma^{\ul a}{}_{c_1\cdots}=\Gamma^{\ul a}{}_{{\ul b}\cdots}
E_{c_1}^{\ul b}\cdots.$

The last term in this equation can be shown to vanish by Fierz
rearranging the expression $\epsilon^{c_1\cdots
c_{p-1}ab}\Gamma_a^{\alpha\beta}\Gamma_b^{\gamma\delta}$, and then using
the following identities (in fact, we need only the trace of
these identities in their symmetrized indices):
$$
\eqalign{
& \Gamma_{(b}^{\alpha'\beta'}
\big (E_{\alpha'r}\eta\Gamma^{\ul{c_1}}{}_{a)}E_{\beta's} \big)=0
\ ,\qquad (p=2;\ n \ne 1,2)\ , \cr
& \big(\Gamma_5\Gamma_{c(b}\big)^{\alpha\beta}\big
(E_\alpha\eta\Gamma^{\ul{c_1}}{}_{a)c_2}E_\beta \big)=0\ ,\qquad
(p=3)  \ , \cr
& \Gamma_{b_1\cdots (b_r}^{\alpha\beta}\big
(E_\alpha\eta\Gamma^{\ul{c_1}}{}_{a)c_2\cdots c_{p-1}}E_\beta \big)=0\ ,
\qquad (p=3,\ r=1,2\ ;\ p=5,\ r=1) \ .\cr }\eqno(3.43)
$$
 The above identities can be
derived by  multiplying  the cyclic identity (3.18) with
$\Gamma_a^{\alpha'\beta'}\Gamma_b^{\gamma'\delta'} \hfill\break
E_{\alpha'r}^{\ul \alpha}E_{\beta's}^{\ul\beta}E_{\gamma'q}^{\ul\gamma}
E_{\delta'q}^{\ul\delta}$
or with
$\Gamma_a^{\alpha\beta}\big(\Gamma_5\Gamma_{bc}\big)^{\gamma\delta}
E_{c_2}^{\ul{c_2}} E_\alpha^{\ul \alpha}
E_\beta^{\ul\beta}E_\gamma^{\ul\gamma} E_\delta^{\ul\delta}$, or with
$\Gamma_a^{\alpha\beta}\Gamma_{b_1\cdots b_r}^{\gamma\delta}
E_{c_2}^{\ul c_2}\cdots E_{c_{p-1}}^{\ul{c_{p-1}}} \hfill\break
 E_\alpha^{\ul\alpha}E_\beta^{\ul\beta}E_\gamma^{\ul\gamma}
E_\delta^{\ul\delta}$, respectively, and then using eq. (3.8).

With the last term vanishing in (3.42), it follows that
\footnote{$^\dagger$}{\tenfoot We are grateful to M. Tonin for explaining to us
the derivation of this identity for the supermembrane.}
$$
Q = ({\rm det}\ g)^{1/2}\ . \eqno(3.44)
$$
Substitution of (3.44) into the action (3.36) yields
the following simple result which is a natural generalization of the massive
superparticle case:
$$
\eqalign{
S=& \int d^{p+1}\sigma d^{mn}\theta P_{\ul a}^\alpha
E_\alpha^{\ul a}  +{(p+1)!\over 2}\int d^{p+1}\sigma
\big(-{\rm det}\ E_m^{~\ul a}
E_n^{~\ul a}\big)^{1/2}\vert_{\theta=0}\cr
 &+\int d^{p+1}\sigma \epsilon^{m_1\cdots m_{p+1}}
\partial_{m_{p+1}}Z^{\ul M_{p+1}}\cdots
 \partial_{m_1}Z^{\ul M_1} B_{\ul{M_1}\cdots
\ul{M_{p+1}}}\vert_{\theta=0}\cr}\ ,\eqno(3.45)
$$
where we have used the constraint (3.7) in manipulations similar to (2.34).

	In summary, our main result for super $p$--branes is the action (3.3)
together with the definitions (2.10) and (3.4) and the constraints (3.1) and
(3.2).  Elimination of the Lagrange multiplier $P^{M_1\cdots M_{p+1}}$
yields the result (3.45). Below we shall comment on various aspects of these
results and we shall discuss a number of open problems.

\bigskip
\noindent{\bf 4. Conclusions}
\bigskip
We have found a twistor-like formulation of a class of
super p-brane theories in which $\kappa$-symmetry is replaced
by worldvolume local supersymmetry. The form of the action (3.45) essentially
coincides with the Nambu-Goto form of the usual super $p$--brane action. The
difference is due to the Lagrange multiplier term. It is not altogether clear
whether the equations of motions are equivalent to those which follow from the
usual super $p$--brane action [21]. For this to happen, one must show that
there is sufficiently powerful gauge symmetry of the action which makes it
possible to gauge away the Lagrange multiplier. We have shown that for the
massive superparticle such a gauge symmetry indeed exists (see eq. (2.37)).
The existence of this gauge symmetry relies on the Dirac matrix identity
(2.32). It remains to be seen whether a similar gauge symmetry exists for
other values of $p$. We expect that the $p$--brane Dirac matrix identity
(3.18) will play an essential role in proving the existence of such a
symmetry.

One of the essential ingredients of the
twistor-like transform is the existence of  a closed super (p+1)-form on the
worldvolume superspace which is constructed out of the pull-backs of a super
$(p+1)$-form and its curvature in target superpspace. We have shown that this
closed $(p+1)$-form exists for the cases $(p,m,n) = (2,2,8), (5,4,2), (2,2,4),
(3,4,1), (2,2,2)$ and $(2,2,1)$. The $p=2$ cases were already considered in
[16]. We believe that the existence of this closed
$(p+1)$-form should have some interesting geometric interpretation, independent
of the role it plays in the twistor-like transform. For instance, it seems
that it is related to the light-like integrability principle
[24][13]. We also note an interesting relation between our work and that of
[25][26]. In both cases the tension parameter is supposed to emerge
as an integration
constant of the equations of motion. The p-form gauge potential
occurring in [26] seems to be closely related to the p-form
gauge potential $Q_{M_1\cdots M_p}$ occurring in our work.
We hope that a more precise understanding of all
these connections may lead to a
better understanding of the theories in question.

There are a number of open problems which deserve futher investigation. To
name a few, what is the
precise relation between  our action and the usual one [21] at the quantum
level? What are the physical degrees of freedom described by this action? Are
the symmetries of the action anomaly-free? Can the quantization problems of the
usual $\kappa$--symmetric action be avoided by the new action? Is the theory
finite?

       Anothere open problem of considerable interest is how to couple
Yang-Mills sector to the theory (such theories are usually referred to as
heterotic $p$--brane theories, because of their similarity to the heterotic
string theory). It is tempting to think that since in the twistor-like
formulation the local worldvolume supersymmetry is manifest in a superspace
formalism, one may simply use the body of knowledge available on superspace
formulation of matter/Yang-Mills systems coupled to supergravity. However,
there is an unusual property of the twistor-like formulations, namely, the
local supersymmetry does not seem to require kinetic terms for the
supergravity multiplet. On the other hand, in a supergravity plus
matter/Yang-Mills system, typically one encounters these kinetic terms. Thus,
one may look for different than usual local supersymmetric invariants (using
the usual kind of tensor calculus when available) or consider the possibility
of including the supergravity kinetic terms in the spirit of ref. [28], where
such terms do arise in the context of finding effective actions for heterotic
$p$--brane solitons. We hope that the results of this paper will help in the
eventual solution of this problem.

 \np
\centerline{\bf Acknowledgements}
\bigskip
E.S.~would like to thank Nathan Berkovits and Kelly Stelle for stimulating
discussions, and Imperial College, Groningen University and the International
Center for Theoretical Physics in Trieste for hospitality. We gratefully
acknowledge very helpful communications with Mario Tonin.
The work of E.B.~has been made possible by a fellowship of the Royal
Netherlands Academy of Arts and Sciences (KNAW).

\np
\vskip 1.5truecm
\centerline{\bf References}
\bigskip
\item{1.} W. Siegel, Phys. Lett. {\bf B128} (1983) 397; Class. Quantum
          Grav. {\bf 2} (1985) 195.
\item{} M.B. Green and J.H. Schwarz, Phys. Lett. {\bf B136 } (1984) 367.
\item{2.} D.P. Sorokin, V.I. Tkatch and D.V. Volkov, Mod. Phys. Lett. {\bf A4}
         (1989) 901;
\item{}   D.P. Sorokin, V.I. Tkacth, D.V. Volkov and A.A. Zheltukhin,
          Phys. Lett. {\bf B216} (1989) 302.
\item{3.} N. Berkovits, Phys. Lett. {\bf B232} (1989) 184;
 Phys. Lett. {\bf B300} (1992) 53; Nucl. Phys. {\bf B379} (1992) 96.
\item{4.} I. Bengston and M. Cederwall, Nucl. Phys. {\bf B302} (1988) 81.
\item{5.} N. Berkovits, Phys. Lett. {\bf B247} (1990) 45.
\item{6.} F. Delduc and E. Sokatchev, Class. Quantum. Grav. {\bf 9} (1992)
         361.
\item{7.} P.S. Howe and P.K. Townsend, Phys. Lett. {\bf B259} (1991) 285.
\item{8.} A.S. Galperin, P.S. Howe and K.S. Stelle, Nucl. Phys. {\bf B368}
          (1992) 281.
\item{9.} P.S. Howe and P.C. West, Int. J. Mod. Phys. {\bf A7} (1992) 6639.
\item{10.} A. Galperin and E. Sokatchev, Phys. Rev. {\bf D46} (1992) 714.
              175.
\item{11.} N. Berkovits, Phys. Lett. {\bf B232} (1989) 184;
\item{} F. Delduc, E. Ivanov and E. Sokatchev, Nucl. Phys. {\bf B384} (1992)
        334.
\item{12.} M. Tonin, Phys. Lett. {\bf B266} (1991) 312; Int. J. Mod. Phys.
            {\bf A7} (1992) 6013;
\item{} N. Berkovits, Nucl. Phys. {\bf B379} (1992) 96;
\item{13.} E. Delduc, A. Galperin, P.S. Howe and E. Sokatchev, Phys. Rev.
          {\bf D47} (1993) 578.
\item{14.} D.P.~Sorokin and M.~Tonin, {\it On the chiral fermions
          in the twistor-like formulation of $D=10$ heterotic string},
          preprint, DFPD/93/TH/52 (July 1993).
\item{15.} A.~Galperin and E.~Sokatchev, {\it A twistor formulation of the
          non-heterotic superstring with manifest worldsheet supersymmetry},
          preprint, BONN-HE-93-05 (April 1993).
\item{16.} P. Pasti and M. Tonin, {\it Twistor--like formulation of the
          supermembrane in D=11}, preprint, DFPD/93/TH/07 (February 1993).
\item{17.} E.A. Ivanov and A.A. Kapustnikov, Phys. Lett. {\bf B267} (1991)
\item{18.} J.P. Gauntlett, Phys. Lett. {\bf B272} (1991) 25.
\item{19.} E. Sezgin, {\it Aspects of $\kappa$--symmetry}, preprint, CTP
           TAMU-28/93 (October 1993).
\item{20.} J. Hughes, J. Liu and Polchinski, Phys. Lett. {\bf B180} (1986)
370.
\item{21.} E. Bergshoeff, E. Sezgin and P.K. Townsend, Phys. Lett. {\bf B189}
(1987) 75.
\item{22.} A.~Ach\'ucarro, J.M.~Evans, P.K.~Townsend and D.L.~Wiltshire, Phys.
Lett. {\bf B198} (1987) 441.
\item{23.} E.~Cremmer and S.~Ferrara, Phys.~Lett.B91 (1980) 61;
\item{} L.~Brink and P.S.~Howe, Phys.~Lett.~B91 (1980) 384.
\item{24.} E. Witten, Nucl. Phys. {\bf B266} (1986) 241;
\item{} L. Chau and B. Milewski, Phys. Lett. {\bf B216} (1989) 330;
\item{} E. Bergshoeff, P.S. Howe, C.N. Pope, E. Sezgin and E. Sokatchev,
        Nucl. Phys. {\bf B354} (1991) 113.
\item{25.} J.A.~de Azc\'arraga, J.M.~Izquierdo and P.K.~Townsend,
Phys.~Rev.~D45 (1992) 3321;
\item{} P.K.~Townsend, Phys.~Lett.~277B (1992) 285.
\item{26.} E.~Bergshoeff, L.A.J.~London and P.K.~Townsend,
           Class.~Quantum grav.~9 (1992) 2545.
\item{27.} M. Tonin, private communication.
\item{28.} A. Strominger, Nucl. Phys. {\bf B343} (1990) 167. [Erratum:
            {\bf B353} (1991) 565].

\np

$$
\vbox{\def\tablerule{\noalign{\hrule}}
\offinterlineskip\baselineskip=16pt\halign{\strut#&
\vrule#&
\hfil#\hfil&\vrule#&\hfil#\hfil&\vrule#&\hfil#\hfil&\vrule#&
\hfil#\hfil&\vrule#&\hfil#\hfil&\vrule#&\hfil#\hfil&\vrule#&
\hfil#\hfil&\vrule#&\hfil#\hfil&\vrule#&\hfil#\hfil&\vrule#&
\hfil#\hfil&\vrule#&\vrule#\cr
\tablerule
     &&&\omit&&\omit&&\omit&&\omit&&\omit&&\omit&&\omit
     &&\omit&&\cr
&&\multispan{19}\hfil Target Space Data \hfil &\cr
     &&&\omit&&\omit&&\omit&&\omit&&\omit&&\omit&&\omit&&\omit
     &&\omit&&\cr
\tablerule
&&$d$&&$11$&&$10$&&$9$&&$8$&&$7$&&$6$&&$5$&&$4$&\cr
\tablerule
&&\ $(M,N)\ $&&(32,1)&&(16,1)&&(16,1)&&(16,1)&&(8,2)
&&(4,2)&&(4,2)&&(4,1)&\cr
\tablerule
&&${\ul G}$&&--&&--&&--&&--&&\ USp(2)\ &&\ USp(4)\ &&\ USp(4)\
&&\ SO(4)\ &\cr
\tablerule
&&$C_{\ul{\alpha'\beta'}}$&&A&&A&&S&&S&&S&&S&&S&&A&\cr
\tablerule
&&$\Gamma^{\ul a}_{\ul{\alpha'\beta'}}$&&S&&S&&S&&S &&A&&A
&&A&&S&\cr
\tablerule
&&$\eta_{\ul{rs}}$&&--&&--&&--&&--&&A&&A&&A&&S&\cr
\tablerule
&&Type&&M&&MW&&PM&&PM&&SM&&SMW&&SM&&M&\cr
\tablerule
 &&&\omit&&\omit&&\omit&&\omit&&\omit&&\omit&&\omit
    &&\omit&&\omit&&\cr
&&\multispan{19}\hfil Worldvolume  Data ($p\ge 2$) \hfil&\cr
    &&&\omit&&\omit&&\omit&&\omit&&\omit&&\omit&&\omit
    &&\omit&&\omit&&\cr
\tablerule
&&$p$&&$2$&&$5$&&$4$&&$3$&&$2$&&$3$&&$2$&&$2$&\cr
\tablerule
&&$(m,n)$&&(2,8)&&(4,2)&&(4,2)&&(4,2)&&(2,4)&&(4,1)
&&(2,2)&&(2,1)&\cr
\tablerule
&&$G$&&\ SO(8)\ &&\ USp(2)\ &&\ USp(2)\
&&\ SO(2)\ &&SO(4)&&--&&SO(2)&&-- &\cr
\tablerule
&&$C_{\alpha'\beta'}$&&A&&S&&S&&A&&A&&A&&A&&A&\cr
\tablerule
&&$\Gamma^a_{\alpha'\beta'}$&&S&&A&&A&&S&&S&&S&&S&&S&\cr
\tablerule
&&$\eta_{rs}$ &&S&&A&&A&&S&&S&&--&&S&&--&\cr
\tablerule
&&Type&&M&&SMW&&SM&&M&&M&&M&&M&&M&\cr
\tablerule}}
$$

\vskip 1.5truecm
Table 1.  In this table $d$ indicates the
dimension of spacetime, $M$ is the dimension of the spinor irrep of
$SO(d-1,1)$, $N$ is the dimension of the defining
representation of the automorphism group ${\ul G}$ of the super Poincar\'e
algebra in $d$ dimensions, $C_{\ul{\alpha'\beta'}}$ is the charge
conjugation matrix, $\Gamma^{\ul a}_{\ul{\alpha'\beta'}}$ are the Dirac
matrices $(\Gamma^{\ul a}C)_{\ul{\alpha'\beta'}}$and $\eta_{\ul{rs}}$ is the
invariant tensor of ${\ul G}$. We often use the notation in which a pair of
indices $({\ul \alpha'}r)$ is replaced by a single index ${\ul \alpha}$.
Furthermore, in $d=6,10$ the matrices $\Gamma^{\ul a}_{\ul {\alpha\beta}}$ are
chirally projected  Dirac matrices and $\Gamma_{\ul a}^{\ul{\alpha\beta}}$
are projected with opposite chirality. In this notation
raising or lowering of the spinor indices is not needed. The types of spinors
are characterized according to the reality and chirality conditions imposed on
them, namely Majorana (M), pseudo-Majorana (PM), symplectic Majorana (SM),
Majorana-Weyl (MW) and symplectic Majorana-Weyl (SMW). Corresponding
quantities are listed for the super $p$--branes that arise in target space
dimension $d$.

\end